\documentclass[11pt]{article}
\usepackage[latin1]{inputenc}
\usepackage[T1]{fontenc}
\usepackage{amsmath}
\usepackage{amsfonts}
\usepackage{amssymb}
\usepackage{color}
\usepackage{graphicx}
\title{Functional bosonization of a Dirac field in $2+1$ dimensions, in the
presence of a boundary}
\author{C.~D.~Fosco$^a$   and
F.~A.~Schaposnik$^b$
\\
~
\\
~
\\
{\normalsize $^a\!$\it Centro At\'omico Bariloche and Instituto
Balseiro,}\\
{\normalsize $\!$\it Comisi\'on Nacional de Energ\'\i a At\'omica, 8400
Bariloche, Argentina}\\
{\normalsize $\!$\it }\\
{\normalsize $^b\!$\it Departamento de F\'\i sica, Universidad
Nacional de La Plata}\\ {\normalsize\it Instituto de F\'\i sica La Plata-CONICET}\\
{\normalsize\it C.C. 67, 1900 La Plata,
Argentina}
 }
\begin{document}
\date{\today}
\maketitle
\begin{abstract}
We apply the functional bosonization procedure to a massive Dirac field
defined on a $2+1$ dimensional spacetime which has a non-trivial boundary.
We find the form of the bosonized current both for the bulk and boundary
modes, showing that the gauge field in the bosonized theory contains a
perfect-conductor boundary condition on the worldsheet spanned by the
boundary. We find the bononized action for the corresponding boundary modes.
\end{abstract}
\section{Introduction}
A seemingly obvious yet fruitful property of quantum field theory systems
is that they must be susceptible of being described in terms of different
sets of fields. This finds an extreme realization in the
bosonization  procedure, whereby a model can be defined
in terms of either fermionic or bosonic quantum fields, the equivalence
between those two formulations is made explicit by the existence of so
called `bosonization rules'. Besides mapping one set of fields into the
other, they yield the dynamics the new variables are subjected to, in order
to correspond to the same physical model.

In $1+1$ spacetime dimensions, bosonization is a very powerful tool which
allows one to understand, and in some cases even to solve, some non-trivial
Quantum Field Theory models (see  \cite{Stone:1995ys} for a complete list of references). It is interesting to note that there is no fundamental theoretical stumbling block to the extension of this path-integral approach to higher dimensions.  Indeed, there has been some progress in the
application, although in an approximated form, of a path integral
bosonization procedure to theories in more than two spacetime dimensions,
dealing with both the Abelian and the non-Abelian cases.

We are concerned here with $2+1$ spacetime dimensions, where the path integral bosonization framework yields the exact form of the bosonized form of the current, while an inverse mass expansion can been used to determine the corresponding local terms in the dual bosonic action. Locality plus gauge invariance strongly constraint the form of the possible terms. Indeed, the leading term in the dual action becomes a Chern-Simons term, while the next-to-leading one corresponds, in the Abelian or non Abelian cases, to a (local) Maxwell \cite{FS}-\cite{c}  or Yang-Mills term \cite{b}, respectively.
 When the fermions are massless, the above procedure becomes more involved, since the even parity part of the dual action becomes non local,  involving the squared root of the Laplacian \cite{c}.  Note, however, that the bosonization rule for the current is still the same as in the massive
case, and that the dual action still contains a Chern-Simons term. The need for the latter has been shown explicitly, as a consequence of an $eta$ function regularization required to have a consistent gauge invariant theory \cite{SSWW}.

Let us finally point out  that the path integral bosonization approach  can be also  employed in higher dimensions, and to situations where the fermionic theory has more than one conserved currents. For example, bosonization rules for   fermionic currents in $3+1$ space-time dimensions have been found in terms of Kalb-Rammond fields  \cite{Fo}.

In this paper, we are concerned with massive (${\rm mass}\equiv m$)
fermions on a $2+1$ dimensional spacetime with a non-trivial boundary
We are concerned here with massive adapting those path-integral bosonization results in three space-time dimensions to a situation
where there is a non-trivial boundary. Besides dealing with the necessary
changes one has to implement to cope with it (a non-trivial boundary calls
for a non-trivial boundary condition), we also include an auxiliary source
for the fermionic current, localized on the boundary of the spacetime
manifold. This last step will allow us, as we shall see, to express the
current corresponding to the boundary modes in terms of the (bulk)
bosonized current.

This paper is organized as follows: in Sect.~\ref{sec:derivation} we
present the derivation of the bosonized version of the model, within the
context of the path integral formulation.
Then, in Sect.~\ref{sec:discussion}, we study the properties of the
resulting bosonic theory, and present our conclusions.
\section{Generating functional}\label{sec:derivation}
To begin with, let us introduce ${\mathcal S}_f({\bar\psi},\psi)$, the
Euclidean action for a (free) massive Dirac field in $2+1$ dimensions:
\begin{equation}\label{eq:defsf}
{\mathcal S}_f({\bar \psi},\psi) \;=\; \int d^3x \,
{\bar \psi} (\not \! \partial + m) \psi \;,
\end{equation}
where, for Dirac's $\gamma$-matrices, we have adopted the conventions:
\begin{equation}\label{eq:conv}
	\gamma_\mu^\dagger \,=\, \gamma_\mu \;,\;\; \{ \gamma_\mu , \gamma_\nu
	\} \,=\, 2 \, \delta_{\mu\nu} \;.
\end{equation}
Letters from the middle of the Greek alphabet are assumed to run over the values
$0,\,1,\,2$. The Euclidean metric has been assumed to be the identity matrix
$\delta_{\mu\nu}$. We shall sometimes raise of lower a spacetime index for
notational convenience, although, for this metric tensor, there is no difference
between them.

To proceed, we need to deal with the fermionic current, $J_\mu = {\bar
\psi} \gamma_\mu \psi$. A first step will be to introduce an auxiliary
source $s_\mu$, which will allow us to generate correlation functions
involving that operator, in the same way as when bosonization is
constructed in the no-boundary case. This will amount to adding to the
fermionic action an extra term ${\mathcal S}_J(s,J)$, where
\begin{equation}
	{\mathcal S}_J(s,J) \;=\; i \, \int d^3x  \, s_\mu(x) J_\mu(x) \;.
\end{equation}

The current appears also as part of a constraint, namely, that its normal component, $J_n$, vanishes on
${\mathcal M} = \partial U$, the boundary of $U$, the spacetime region the
field is confined to. The vanishing of the normal component of the current
ensures that the fermions are indeed confined to $U$.  Let us now introduce
an explicit form for that constraint. To that end, we assume that a
parametrization has been introduced for ${\mathcal M}$:
\begin{equation}
	\sigma \,=\, (\sigma^0, \sigma^1) \;\to\; y_\mu(\sigma) \;, \;\;
	\mu \,=\, 0,\,1,\,2 \;,
\end{equation}
with the two parameters $\sigma^\alpha$, $\alpha \,=\, 0,\, 1$.
In terms of the parametrization, the unit normal $\hat{n}_\mu$ may be
written as follows:
\begin{equation}
	\hat{n}_\mu(\sigma) \;=\; \frac{N_\mu(\sigma)}{\sqrt{N^2(\sigma)}}
	\;\;,\;\;\;  N_\mu(\sigma) \,=\, \frac{1}{2}\epsilon^{\alpha\beta}
	\, \epsilon_{\mu\nu\lambda} e^\nu_\alpha(\sigma)
e^\lambda_\beta(\sigma) \;,
\end{equation}
where we have introduced the tangent vectors
$e^\mu_\alpha(\sigma) = \frac{\partial y^\mu}{\partial
\sigma^\alpha}(\sigma)$.

Thus, the constraint can be conveniently introduced in terms of a functional Fourier
representation, at the expense of using an auxiliary scalar field,
$\xi(\sigma)$,
living on ${\mathcal M}$:
\begin{align}\label{eq:defsm}
\delta_{\mathcal M}(J_n) &=\; \int {\mathcal D} \xi \;
e^{- \, S_{\mathcal M}(\xi, J)} \;\;, \nonumber\\
S_{\mathcal M}(\xi, J) &=\; i \, \int d^2\sigma \, \sqrt{g(\sigma)} \,
\xi(\sigma) \, {\hat n}_\mu(\sigma) J_\mu(y(\sigma)) \;,
\end{align}
with $g(\sigma) \equiv \det[g_{\alpha\beta}(\sigma)]$,
$g_{\alpha\beta}(\sigma) \,=\, e^\mu_\alpha(\sigma) e^\mu_\beta(\sigma)$
denoting the induced metric on ${\mathcal M}$.

Therefore, putting together the previous elements, we see that a
generating functional of current correlation functions,
${\mathcal Z}(s)$, for a massive Dirac field in $2+1$ Euclidean dimensions, in the
presence of a boundary ${\mathcal M}$, may be written as follows:
\begin{equation}\label{eq:defzs}
{\mathcal Z}(s) \;=\; \int {\mathcal D}\psi \,{\mathcal D}{\bar
\psi} \; \delta_{\mathcal M}(J_n) \; e^{-{\mathcal S}({\bar \psi},\psi;s)
} \;,
\end{equation}
with
\begin{equation}
{\mathcal S}({\bar \psi},\psi;s) \;=\;{\mathcal S}_f({\bar \psi},\psi)
\,+ \,{\mathcal S}_J(s,J) \;.
\end{equation}
Equivalently, recalling the representation (\ref{eq:defsm}),
\begin{equation}\label{eq:defzss}
{\mathcal Z}(s) \;=\; \int {\mathcal D}\psi \,{\mathcal D}{\bar
\psi} \, {\mathcal D}\xi \; e^{-{\mathcal S}({\bar \psi},\psi;s)
	\,-\,S_{\mathcal M}(\xi, J)} \;.
\end{equation}

The functional ${\mathcal S}_{\mathcal M}$, introduced in (\ref{eq:defsm})
is explicitly reparametrization invariant. Besides, since $\sqrt{N^2(\sigma)} =
\sqrt{g(\sigma)}$, we see that it may be rendered also as follows:
\begin{equation}
S_{\mathcal M}(\xi, J) \;=\; i \int d^2\sigma \,
\xi(\sigma) \, N_\mu(\sigma) J_\mu(y(\sigma)) \;,
\end{equation}
or, more conveniently from the point of view of the next steps in our
derivation, also as:
\begin{equation}
	S_{\mathcal M}(\xi, J) \;=\; i \, \int d^3 x \, c_\mu(x) \, J_\mu(x) \;,
\end{equation}
with:
\begin{equation}\label{eq:defc}
	c_\mu(x) \;=\; \int d^2\sigma \, \xi(\sigma) \, N_\mu(\sigma) \,
\delta[x - y(\sigma)] \;.
\end{equation}

Then, the generating functional may be written as follows:
\begin{equation}
{\mathcal Z}(s) \;=\; \int \, {\mathcal D}\xi \, {\mathcal D}\psi \,{\mathcal D}{\bar
\psi} \; e^{-{\mathcal S}_f({\bar \psi},\psi;s+c)} \;,
\end{equation}
with
\begin{equation}
{\mathcal S}_f({\bar \psi},\psi;s) \;=\; \int d^3x \,
{\bar \psi} (\not \! \partial + i \not \! s + m) \psi \;.
\end{equation}

We then perform the change of variables:
\begin{equation}
	\psi(x) \to e^{i \alpha(x)} \psi(x) \;\;, \;\;\;
	{\bar\psi}(x) \to e^{-i \alpha(x)} {\bar\psi}(x) \;,
\end{equation}
and integrate over $\alpha$, to obtain (discarding immaterial factors)
\begin{equation}\label{eq:zseq1}
{\mathcal Z}(s) \;=\;  \int {\mathcal D} \alpha \;{\mathcal D}\xi
\;{\mathcal D}\psi \,{\mathcal D}{\bar \psi} \;
e^{-{\mathcal S}_f({\bar \psi},\psi;s+c+\partial\alpha)} \;.
\end{equation}
Finally, we make the substitution $\partial_\mu \alpha \to b_\mu$,
\begin{equation}\label{eq:zseq2}
	{\mathcal Z}(s) \;=\;  \int {\mathcal D}b  \; \delta[{\tilde
	f}_\mu(b)] \,
	{\mathcal D}\xi \;{\mathcal D}\psi \,{\mathcal D}{\bar \psi} \;
e^{-{\mathcal S}_f({\bar \psi},\psi;s+c+b)} \;,
\end{equation}
where the condition ${\tilde f}_\mu(b) = \epsilon_{\mu\nu\lambda} \partial_\nu
b_\lambda =0$, which implies that $b_\mu$ is a pure gradient~\footnote{We
assume that $U$ is a simply connected manifold.} has been introduced in
the measure.

Introducing yet another auxiliary field, $A_\mu$, to implement that condition:
\begin{equation}
	\delta[{\tilde f}_\mu(b)] \;=\; \int {\mathcal D}A \; e^{-i \int
	d^3x \, A_\mu {\tilde f}_\mu(b)} \;,
\end{equation}
we get:
\begin{equation}\label{eq:zseq3}
{\mathcal Z}(s) \;=\;  \int {\mathcal D}A \; {\mathcal D}b  \;
	{\mathcal D}\xi \;{\mathcal D}\psi \,{\mathcal D}{\bar \psi} \;
	e^{-{\mathcal S}_f({\bar \psi},\psi;s+c+b) - i \int d^3x A_\mu
	{\tilde f}_\mu(b) } \;.
\end{equation}
Finally, we make the shift $b \to b - c - s$, to obtain:
\begin{equation}\label{eq:zseq4}
{\mathcal Z}(s,t) \;=\;  \int {\mathcal D}A \; {\mathcal D}b  \;
	{\mathcal D}\xi \;
	e^{-W(b) - i \int d^3x A_\mu
	[ {\tilde f}_\mu(b) - {\tilde f}_\mu(c) - {\tilde f}_\mu(s)]} \;,
\end{equation}
where $W(b)$ is the effective action for the $b_\mu$ field due to the
fermion loop, namely,
\begin{equation}\label{eq:defzb}
	e^{-W(b)} \;=\; \det (\not \! \partial + i \not \! b + m) \;.
\end{equation}

The next step is to integrate out the auxiliary fields; to that end, we
first rearrange the integrals as follows:
\begin{align}\label{eq:zseq41}
	{\mathcal Z}(s) &=\;  \int {\mathcal D}A \; e^{i \int d^3x s_\mu {\tilde f}_\mu(A) } \; \nonumber\\ & \times \; \int {\mathcal D}\xi \;
e^{ i \int d^3x c_\mu {\tilde f}_\mu(A) } \;
\Big( \int {\mathcal D}b  \;
e^{-W(b) - i \int d^3x b_\mu  {\tilde f}_\mu(A)} \Big) \;.
\end{align}

The integral over the $b_\mu$-field requires the knowledge of the fermionic
determinant. Assuming that the large-mass expansion is applicable, we have,
keeping the leading term \cite{GRS}:
\begin{equation}
	W(b) \; \simeq \; \pm \frac{i}{4\pi} \, \int d^3x \,
	\epsilon_{\mu\nu\lambda} b_\mu \partial_\nu b_\lambda \;.
\end{equation}

Thus, the integral over $b_\mu$ yields, in this approximation:
\begin{equation}
\int {\mathcal D}b  \;
e^{-W(b) - i \int d^3x b_\mu  {\tilde f}_\mu(A)}  \;=\;
e^{ \pm  \frac{i}{2} \, \int d^3x \, 2 \pi  \,
\epsilon_{\mu\nu\lambda} A_\mu \partial_\nu A_\lambda} \;.
\end{equation}

Performing the rescaling $A_\mu \to \frac{1}{\sqrt{2\pi}} A_\mu$, and
defining
\begin{equation}
	J_\mu \;\to\; \frac{i}{\sqrt{2\pi}}  \epsilon_{\mu\nu\lambda}\partial_\nu A_\lambda
	\;\equiv \; {\mathcal J}_\mu
	\;,
\end{equation}
which is the expression for the bosonized current, as seen by taking the
functional derivative with respect to $s_\mu$, we get:
\begin{align}\label{eq:zseq42}
	{\mathcal Z}(s) &=\;  \int {\mathcal D}A \; e^{\int d^3x s_\mu {\mathcal J}_\mu } \; \nonumber\\
	& \times \; \int {\mathcal D}\xi \;
	e^{\int d^3x c_\mu {\mathcal J}_\mu  \; \pm \; \frac{i}{2}\, \int d^3x \,
\epsilon_{\mu\nu\lambda} A_\mu \partial_\nu A_\lambda} \;.
\end{align}
Or,
\begin{equation}\label{eq:zseq43}
	{\mathcal Z}(s,t) \;=\;  \int {\mathcal D}A \, \delta_{\mathcal
	M}({\mathcal J}_n) \, e^{\int d^3x s_\mu {\mathcal J}_\mu \,\pm \,\frac{i}{2} \int d^3x
\,\epsilon_{\mu\nu\lambda} A_\mu \partial_\nu A_\lambda }\;.
\end{equation}
Which is our final expression for the bosonized version of the system. Note that
the original constraint has been converted into the vanishing of
\mbox{${\mathcal J}_n \equiv \hat{n}_\mu {\mathcal J}_\mu$}, the normal component of the
{\em bosonized\/} current, ${\mathcal J}_\mu$, on ${\mathcal M}$.

Now, regarding $A_\mu$ as an Abelian gauge field, one can show, after
some algebra, that the vanishing of the normal component of the bosonized
current amounts to perfect conductor boundary conditions for that field.
Indeed, the condition:
\begin{equation}
	N^\mu(\sigma) \, {\mathcal J}_\mu(y(\sigma)) \;=\;0
\end{equation}
becomes, in terms of ${\mathcal A}_\alpha(\sigma) \equiv A_\mu(y(\sigma))
e^\mu_\alpha(\sigma)$, the components of $A_\mu(x)$ projected to
${\mathcal M}$,
\begin{equation}
	\partial_\alpha {\mathcal A}_\beta(\sigma) - \partial_\beta
	{\mathcal A}_\alpha(\sigma) \;=\; 0 \;,
\end{equation}
which are perfect-conductor boundary conditions: since the boundary is
two-dimensional, just the vanishing of the
parallel component of the electric field.

A related observation is that one can verify that
\begin{equation}\label{eq:term}
\frac{i}{\sqrt{2\pi}} \int d^3x c_\mu \epsilon_{\mu \nu
\lambda}\partial_\nu A_\lambda \;=\; \frac{i}{\sqrt{2\pi}} \int_{\mathcal
M} d^2 \sigma
\, \xi(\sigma) \, \epsilon^{\alpha\beta} \, \partial_\alpha {\mathcal A}_\beta(\sigma)
\end{equation}
where the rhs depends on the projected components of the gauge field. Now
one can reinterpret the reasoning leading to the perfect-conductor
boundary conditions as follows: the term (\ref{eq:term}), the only place
where the auxiliary field $\xi$ appears, is invariant under constant shifts
of $\xi$: $\xi(\sigma) \to \xi(\sigma) + c_0$. This global continuous
transformation implies, via Noether's theorem, the existence of a conserved
current which is concentrated on the boundary:
\begin{equation}\label{eq:cc1}
	\partial_\alpha j^\alpha(\sigma)  \;=\; 0 \;,\;\; j^\alpha(\sigma)
	\;=\; - \epsilon^{\alpha\beta} \, {\mathcal A}_\beta(\sigma) \;.
\end{equation}

\section{Discussion}\label{sec:discussion}
Let us now consider the evaluation of the constrained path integral
(\ref{eq:zseq43}). The gauge field satisfies
perfect-conductor boundary conditions on the boundary ${\mathcal M}$,
and the exponent  contains a Chern-Simons term, plus terms where the
gauge field couples linearly to sources. We will proceed to split $A_\mu$
in the measure into a classical part $A^{cl}_\mu$, satisfying the proper
boundary conditions, plus a fluctuating field $a^{cl}_\mu$, with trivial (Dirichlet)
boundary conditions:
\begin{equation}\label{eq:split1}
	A_\mu(x) \;=\; 	A^{cl}_\mu(x) \,+\,a_\mu(x) \;,
\end{equation}
such that $a_\mu(x)$ vanishes on ${\mathcal M}$.

Using the definitions: ${\mathcal S}_{CS} \equiv \mp
\frac{i}{2}\, \int d^3x \, \epsilon_{\mu\nu\lambda} A_\mu \partial_\nu
A_\lambda$ and \mbox{$w_\mu \equiv s_\mu + t_\mu$}, the classical equation
of motion have the form:
\begin{equation}\label{eq:split0}
	\delta {\mathcal S}_{CS}(A) \;=\; \delta \big( \int d^3x \, w_\mu
	{\mathcal J}_\mu \big)\;,\;\;\;	
	\partial_\alpha \delta{\mathcal A}_\beta(\sigma) - \partial_\beta
	\delta{\mathcal A}_\alpha(\sigma) \;=\; 0 \;.
\end{equation}
where the last equation follows from the constraint.

The equations above will admit as a solution a sum:
\begin{equation}\label{eq:split2}
	A^{cl}_\mu(x)  \;=\; A^{(0)}_\mu(x) \,+\, A^{(1)}_\mu(x) \,,
\end{equation}
where $A^{(0)}_\mu$ is the general solution to the homogeneous system:
\begin{equation}\label{eq:split3}
	\delta {\mathcal S}_{CS}(A) \;=\; 0  \;,\;\;\;
	\partial_\alpha \delta{\mathcal A}_\beta(\sigma) - \partial_\beta
	\delta{\mathcal A}_\alpha(\sigma) \;=\; 0 \;.
\end{equation}
and  $A^{(1)}_\mu(x)$ a particular solution to the inhomogeneous equation
(i.e., including $w_\mu$).

Let us then consider the equations for $A_\mu^{(0)}$. We see that, because of
the non-trivial boundary, the homogeneous equations are:
\begin{equation}\label{eq:split4}
\int d^3x \, \epsilon_{\mu\nu\lambda} \delta A_\mu  \partial_\nu
A_\lambda \,-\, \frac{1}{2} \int_{\mathcal M}  d^2\sigma \,
\epsilon^{\alpha\beta} {\mathcal A}_\alpha(\sigma)
\delta {\mathcal A}_\beta(\sigma) \;=\;0. \;,
\end{equation}
plus the second equation in (\ref{eq:split0}).

The vanishing of the second term above leaves room for many different
conditions which can be imposed on ${\mathcal A}_\alpha$ to make that
happen. Recalling the conservation of the boundary current
$j^\alpha(\sigma)$ on the boundary, if we want to keep the possibility of
having non-vanishing values for that current, we cannot use trivial
conditions for ${\mathcal A}_\alpha$, since those fields are proportional
to components of the current. In what follows, we assume the border to be
static, namely, to have the form ${\mathcal M} =
{\mathcal C} \times {\mathbb R}$, where ${\mathcal C}$ denotes a static
closed curve: the spatial boundary. Then, for assuming for the current
$j^\alpha$ the form:
\begin{equation}
	j^0(\sigma) \,=\, \rho (\sigma) \;,\;\;\;
	j^1(\sigma) \,=\, \rho (\sigma) \, v
\end{equation}
where $v$ is a constant  with dimensions of velocity, we see that the assumption above
implies, from the continuity equation for the current:
\begin{equation}\label{eq:gc}
	{\mathcal A}_0 \,-\, v {\mathcal A}_1 \;=\; 0 \;.
\end{equation}
The second term in (\ref{eq:split4}) then vanishes; indeed, one first
deduces that $\delta {\mathcal A}_\alpha = \partial_\alpha \omega$, and
then one uses the continuity equation for the surface current.

Then, the rest of the construction is rather standard \cite{Wen:1992vi:}
using general coordinates (rather than Cartesian ones) $x_1$ and $x_2$, such
that the curve ${\mathcal C}$ may be regarded as the coordinate
curve $x_2 = 0$, there are
new coordinates $x'_0 = x_0$, $x'_1 = x_1 + v x_0$ and $x'_2 = x_2$,
such that (\ref{eq:gc}) becomes:
\begin{equation}\label{eq:gc1}
	{\mathcal A}'_0 = 0 \;,
\end{equation}
where ${\mathcal A}'_0$ is the gauge field component in the new
coordinates.

The other two components are pure gauges, and can be extended to pure
gauges over $U$, because of the equations following from the bulk part of
the variation: $A_i = \partial_i \phi$.
Using the independence of the action on the metric, and
extending (\ref{eq:gc1}) to all the spacetime region, as $A'_0(x') =0$,
we see that the action evaluated on this configuration yields:
\begin{equation}
	{\mathcal A}_{CS}(A^{(0)}) \;=\; \pm \frac{i}{2}
	\int d^3x'  \epsilon^{ij} A'_i \partial_0 A'_j
	\;=\; \pm \frac{i}{2}
	\int d^3x'  \epsilon^{ij} { \partial\,}'_{\!i} \phi \,  \partial_0
	{\partial\,}'_{\!j}\phi \;
\end{equation}
or, recalling that the boundary is at $x'_2 = 0$, the action adopts the
Floreanini-Jackiw \cite{FJ} form:
\begin{equation}
{\mathcal A}_{CS}(A^{(0)}) \;=\; \pm \frac{i}{2}
\int dx_0 dx'_1 \big[
\partial_0 \varphi \partial_1 \varphi
- v (\partial_1 \varphi)^2
\big]
\end{equation}
where $\varphi(x_0,x_1) \equiv \phi(x_0,x_1,0)$.

Thus, the classical gauge field configurations contain the $\varphi$ modes,
concentrated on the boundary, which have to be integrated alongside the
fluctuating part $a_\mu$ which has trivial boundary conditions.

The inhomogeneous equation can then be solved by imposing trivial boundary
conditions on $A_\mu^{(1)}$. It is straightforward to see that the
resulting equations and their solutions do not involve the boundary modes.
Finally, the fluctuating part $a_\mu$ appears quadratically and does not
involve the source $s_\mu$, so we can discard it.

Let us end this work by noting that in the last three years there has been
much interest in the application of dualities  to analyze  condensed matter
systems like topological insulators, superconductors, and fractional
quantum  Hall effect systems \cite{SSWW},\cite{w1}-\cite{Witten:2015aoa}. In these studies
bosonization in $2+1$ dimensions play a relevant role \cite{c21}-\cite{c22}
and, in this context, the case of manifolds with boundary like  those we
discussed here would be of interest. We expect to discuss this issue  in a
future publication.

\section*{Acknowledgments}
This research was supported by ANPCyT, CONICET,  UNCuyo and UNLP.


\end{document}